 \documentclass[smallabstract,smallcaptions]{dccpaper}

\usepackage{epsfig}
\usepackage{citesort}
\usepackage{amsmath}
\usepackage{amssymb}
\usepackage{color}
\usepackage{url}
\usepackage{booktabs}

\usepackage{enumitem}
\usepackage{cite}
\usepackage{multirow}
\usepackage{graphicx}
\usepackage{makecell}

\newlength{\figurewidth}
\newlength{\smallfigurewidth}

\begin{document}

\title
{\large
\textbf{Hybrid Local-Global Context Learning for Neural \\Video Compression}
}
\author{%
Yongqi Zhai$^{1,2}$, Jiayu Yang$^{1}$, Wei Jiang$^{1}$, Chunhui Yang$^{1}$,\\ Luyang Tang$^{1,2}$ and Ronggang Wang$^{1,2,3\ast}$\\[0.5em]
\thanks{$\ast$ Ronggang Wang is the corresponding author.}
{\small\begin{minipage}{\linewidth}\begin{center}
\begin{tabular}{c}
$^{1}$Shenzhen Graduate School, Peking University, China \\
$^{2}$Peng Cheng Laboratory, Shenzhen, China \\
$^{3}$Migu Culture Technology Co., Ltd, China \\
\url{zhaiyongqi@stu.pku.edu.cn,}
\thinspace
\url{rgwang@pkusz.edu.cn}
\end{tabular}
\end{center}\end{minipage}}
}

\maketitle
\thispagestyle{empty}

\begin{abstract}
In neural video codecs, current state-of-the-art methods typically adopt multi-scale motion compensation to handle diverse motions. These methods estimate and compress either optical flow or deformable offsets to reduce inter-frame redundancy. However, flow-based methods often suffer from inaccurate motion estimation in complicated scenes. Deformable convolution-based methods are more robust but have a higher bit cost for motion coding. In this paper, we propose a hybrid context generation module, which combines the advantages of the above methods in an optimal way and achieves accurate compensation at a low bit cost. Specifically, considering the characteristics of features at different scales, we adopt flow-guided deformable compensation at largest-scale to produce accurate alignment in detailed regions. For smaller-scale features, we perform flow-based warping to save the bit cost for motion coding. Furthermore, we design a local-global context enhancement module to fully explore the local-global information of previous reconstructed signals. Experimental results demonstrate that our proposed Hybrid Local-Global Context learning (HLGC) method can significantly enhance the state-of-the-art methods on standard test datasets.
\end{abstract}

\Section{Introduction}

Video compression is a fundamental low-level vision task, which aims to reduce the transmission and storage costs of video data. In the past years, neural video compression methods have achieved remarkable progress \cite{lu2019dvc, agustsson2020scale, liu2020neural, hu2021fvc, li2021deep, sheng2022temporal, li2022hybrid, hu2022coarse, yilmaz2023multi, li2023neural, qi2023motion}, and some recent works \cite{li2022hybrid, li2023neural, qi2023motion} even exhibit competitive rate-distortion (RD) performance compared to the latest standard H.266/VVC \cite{bross2021overview}. Most existing neural video compression methods rely on extracting and transmitting inter-frame motion to effectively remove temporal redundancy. Multi-scale motion compensation is widely used in current state-of-the-art methods \cite{sheng2022temporal, li2022hybrid, hu2022coarse, yilmaz2023multi, li2023neural, qi2023motion} to handle diverse motions. According to the information type of motion coding, these methods can be roughly divided into two categories: 1) flow-based methods and 2) deformable convolution-based methods.


Flow-based methods first estimate optical flow at pixel level and then warp the previously reconstructed signals to the target frame for inter-frame prediction. The pioneering DVC \cite{lu2019dvc} used optical flow estimation to replace the block-based motion estimation and performed pixel-level motion compensation. The later work SSF \cite{agustsson2020scale} proposed scale-space flow to reduce the residuals in fast motion area. DCVC \cite{li2021deep} and its following works \cite{sheng2022temporal, li2022hybrid, li2023neural, qi2023motion} were conditional coding frameworks, which warped the previously decoded feature based on optical flow to generate valuable temporal contexts. However, optical flow is difficult to estimate in complex and irregular real-world scenes, especially for regions suffering from occlusion and blur.

Recently, deformable convolution networks (DCN) \cite{dai2017deformable} have been applied in video compression frameworks to achieve better alignment. These methods performed motion estimation and compensation in feature space and compressed the deformable offsets. FVC \cite{hu2021fvc} first employed deformable compensation to replace flow-based warping. Other works used coarse-to-fine motion compensation \cite{hu2022coarse} or multi-scale deformable alignment \cite{yilmaz2023multi} to further improve performance. The increased degree of freedom makes it more robust than optical flow in handling complex motions, but also increases the bit cost for motion coding. Moreover, the training of deformable compensation is unstable without appropriate guidance, which degrades its performance.

For multi-scale compensation frameworks, features at different scales have different characteristics. For example, smaller-scale features mainly focus on large motions, while larger-scale features pay more attention to textures. Taking this into account, we propose a hybrid context generation method that applies different compensation strategies at different scales. Specifically, for the smallest-scale and middle-scale reference features, we perform flow-based warping to save the bit cost for motion coding. For the largest-scale reference feature, with the guidance of optical flow, we stably estimate and compress extra deformable offsets to achieve more accurate compensation in detailed regions. In this way, our hybrid context generation method achieves better RD trade-off between compensation accuracy and bit cost for motion coding.

In addition, existing context enhancement methods \cite{hu2021fvc, hu2022coarse, li2023neural} mainly focus on local inter-frame information and lack the ability to model the long-range correspondence. To this end, we propose a local-global context enhancement module to further boost performance without consuming any bitrate. Specifically, since DCN mainly focuses on local areas, we adopt multi-scale deformable alignment on the generated contexts to reconstruct more accurate details. The estimation of offsets at each scale is guided by smaller-scale offsets and optical flow, which stabilizes training and improves estimation accuracy. Meanwhile, we further design a cross-attention-based enhancement module to extract the global information between frames. Finally, a channel-spatial fusion module is designed to fuse the local and global contexts, which adopts the channel-spatial attention mechanism. Our contributions are summarized as follows:

\begin{itemize}[topsep=2pt, itemsep=2pt, parsep=1pt]
\item[\small\textbullet] We propose a hybrid context generation method for multi-scale motion compensation frameworks, which optimally combines the advantages of flow-based warping and deformable compensation. The proposed method achieves better RD trade-off between compensation accuracy and bit cost for motion coding.
\item[\small\textbullet] We propose a local-global context enhancement module to further enhance the quality of contexts, which utilizes both the local modeling ability of DCN and the global focusing ability of cross-attention mechanism.
\item[\small\textbullet] Experimental results show that our proposed HLGC method can significantly enhance the state-of-the-art methods TCM \cite{sheng2022temporal} and HEM \cite{li2022hybrid}, achieving 16.1$\%$ and 9.1$\%$ bitrate saving in terms of PSNR, respectively.
\end{itemize}

\begin{figure*}[!t]
  \centering
  \includegraphics[width=\linewidth, trim = {172 142 172 142}, clip]{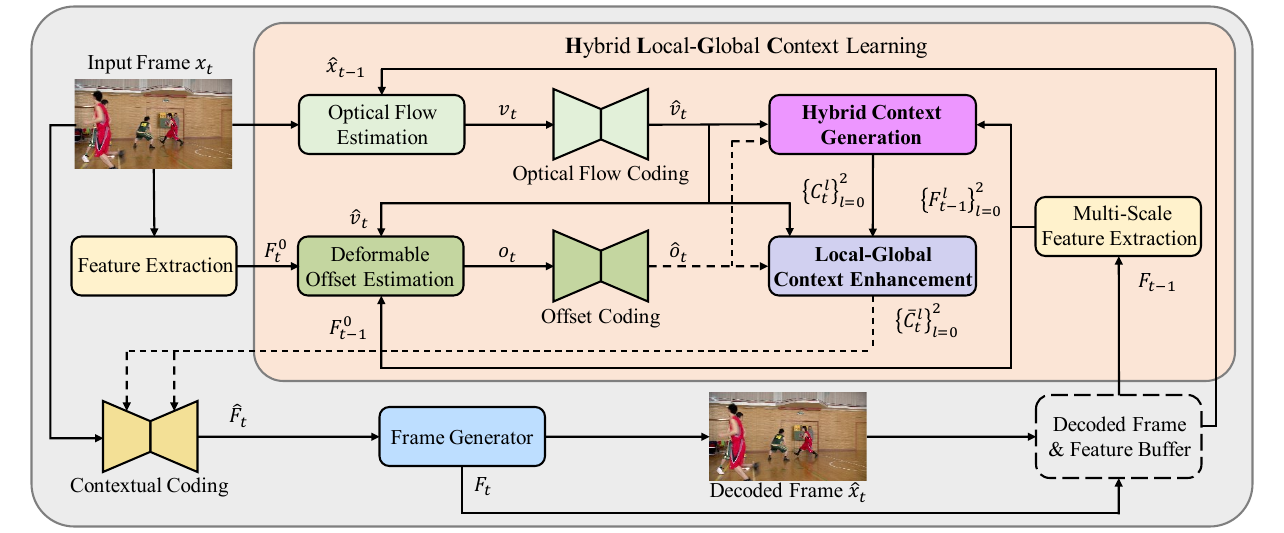}
  \vspace{-1.2cm}
  \caption{The overview of the proposed Hybrid Local-Global Context learning method.}
  \label{overall_framework}
\vspace{-0.8cm}
\end{figure*}

\Section{PROPOSED METHOD}
\SubSection{Overview}

Our proposed HLGC method is integrated into the widely acknowledged baseline TCM \cite{sheng2022temporal} and extended to HEM \cite{li2022hybrid} in the experiments part. Figure~\ref{overall_framework} shows the overview of our HLGC method. In general, HLGC consists of two parts: hybrid context generation and local-global context enhancement. At first, we estimate and compress the optical flow $v_t$ between the input frame $x_t$ and the previous decoded frame $\hat{x}_{t-1}$. With the guidance of the decoded optical flow $\hat{v}_t$, we estimate the extra deformable offsets $o_t$ between current feature $F^0_t$ and reference feature $F^0_{t-1}$. Taking the decoded optical flow $\hat{v}_t$, decoded offsets $\hat{o}_t$, and multi-scale reference features $\{{F}_{t-1}^l\}_{l=0}^2$ as inputs, the hybrid context generation module generates the hybrid temporal contexts $\{{C}_t^l\}_{l=0}^2$. Then, with the assistance of $\hat{v}_t$, $\hat{o}_t$ and $\{{F}_{t-1}^l\}_{l=0}^2$, the local-global context enhancement module further enhances the quality of generated contexts to $\{\bar{C}_t^l\}_{l=0}^2$. Finally, multi-scale enhanced contexts $\{\bar{C}_t^l\}_{l=0}^2$ are used for both contextual encoding and decoding. The proposed modules hybrid context generation and local-global context enhancement are presented in detail in the following subsections.

\SubSection{Hybrid Context Generation}
In \cite{liu2020neural, sheng2022temporal, hu2022coarse}, multi-scale motion compensation has been shown to achieve better alignment results than single-scale method. As for motion coding, existing multi-scale motion compensation methods typically compress either single-scale optical flow \cite{sheng2022temporal, li2022hybrid, li2023neural, qi2023motion} or muti-scale deformable offsets \cite{hu2022coarse, yilmaz2023multi}. However, both methods have their shortcomings. For flow-based warping methods, corresponding downsampled versions of optical flow are used to warp features at different scales. However, it is difficult to handle complex scenes using only flow-based warping. Muti-scale deformable compensation methods are more robust than flow-based warping methods, but require estimating and compressing deformable offsets at each scale, which greatly increases the bit cost for motion coding. Meanwhile, the training of deformable compensation is unstable without appropriate guidance.

\begin{figure*}[!t]
  \centering
  \includegraphics[width=0.7\linewidth, trim = {265 150 245 152}, clip]{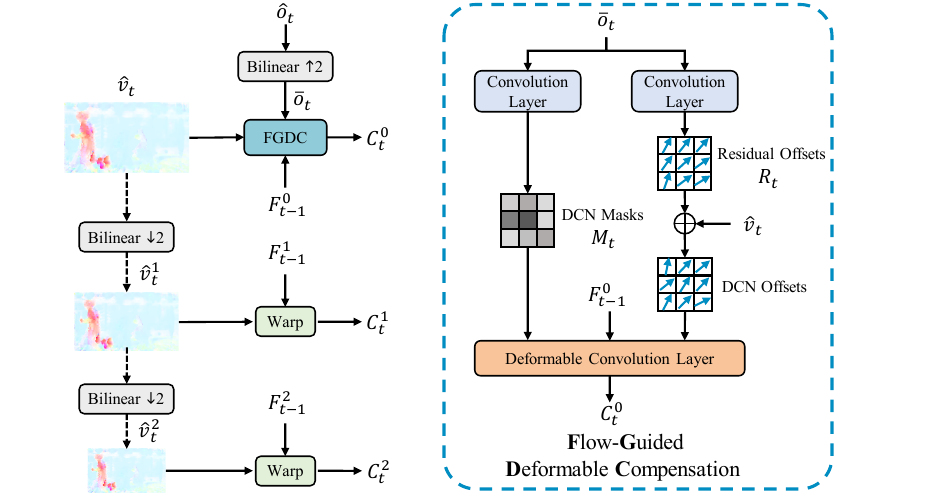}
  \vspace{-0.6cm}
  \caption{ Illustration for the hybrid
context generation module.}
  \label{context-generation}
  \vspace{-0.8cm}
\end{figure*}

The previous work TCM \cite{sheng2022temporal} observed that contexts at different scales have different characteristics. For example, smaller-scale features mainly focus on the regions with large motions, and larger-scale features focus on texture and color information. For video compression task, it is crucial to improve the prediction accuracy while minimizing the bit cost for motion coding. Therefore, in order to obtain better RD performance for inter-frame prediction, we apply different compensation strategies to features at different scales.

As shown in Figure~\ref{context-generation}, our hybrid context generation module combines flow-based warping and deformable compensation. For the middle-scale and smallest-scale features $F^1_{t-1}$ and $F^2_{t-1}$, we apply flow-based warping to save the bit cost for motion coding. Based on the downsampled version of optical flow $\hat{v}^1_{t}$ and $\hat{v}^2_{t}$, we generate the middle-scale and smallest-scale contexts $C^1_{t}$, $C^2_{t}$:
\begin{equation}
\begin{split}
    C^1_{t} = \mathcal{W}(F^1_{t-1}, \hat{v}^1_{t}), 
    \\
    C^2_{t} = \mathcal{W}(F^2_{t-1}, \hat{v}^2_{t}),
\end{split}
\end{equation}
where $\mathcal{W}$ denotes the flow-based warping operator. The largest-scale feature mainly contains detailed information that is critical to the final reconstruction and therefore requires high-accuracy prediction. As shown in Figure~\ref{overall_framework}, to get more accurate alignment in detailed regions, we estimate extra deformable offsets for the largest-scale feature $F^0_{t-1}$ with the guidance of the decoded optical flow $\hat{v}_{t}$. Specifically, we first warp $F^0_{t-1}$ based on $\hat{v}_{t}$ to generate the intermediate predicted feature $\bar{F}^0_{t-1}$:
\begin{equation}
    \bar{F}^0_{t-1} = \mathcal{W}(F^0_{t-1}, \hat{v}_{t}).
\end{equation}
Then, take $F^0_t$, $\bar{F}^0_{t-1}$ and $\hat{v}_{t}$ as inputs, we estimate the refined offsets ${o}_t$:
\begin{equation}
    {o}_t = \mathit{Conv}(F^0_t, \bar{F}^0_{t-1}, \hat{v}_{t}),
\end{equation}
where $\mathit{Conv}$ represents some convolution layers. To reduce the memory cost, ${o}_t$ is estimated to be half the resolution of $F^0_{t-1}$. After offset coding and bilinear upsampling, ${o}_t$ is restored to the original resolution and reconstructed as $\bar{o}_t$. Figure~\ref{context-generation} shows the process of flow-guided deformable compensation (FGDC) operation. Based on $\hat{v}_t$ and $\bar{o}_t$, we perform FGDC on $F^0_{t-1}$ to generate more accurate context $C^0_{t}$, which can be formulated as:
\begin{equation}
    C^0_{t} = \mathit{FGDC}(F^0_{t-1}, \bar{o}_t, \hat{v}_t).
\end{equation}

By combining the advantages of flow-based warping and deformable compensation in an optimal way, our method achieves accurate prediction while reducing the bit cost for motion coding, thereby improving RD performance. The experiments section compares the performance of different compensation strategies at different scales and demonstrate the effectiveness of our hybrid context generation method.

\begin{figure*}[!t]
  \centering
  \includegraphics[width=\linewidth, trim = {140 140 145 140}, clip]{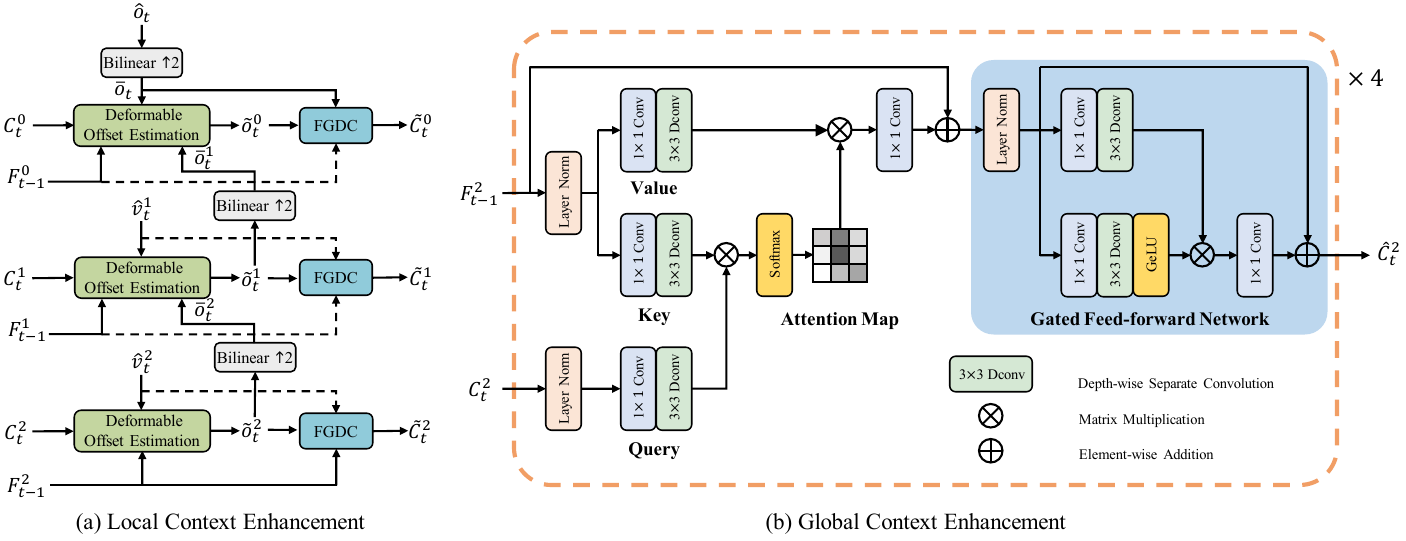}
  \vspace{-1.2cm}
  \caption{Our proposed local-global context enhancement module.}
  \label{context-enhance}
  \vspace{-0.8cm}
\end{figure*}


\SubSection{Local-Global Context Enhancement}
After generating the temporal contexts, previous works proposed many context enhancement methods to further improve the quality of contexts without consuming any bitrate. \cite{hu2021fvc} and \cite{hu2022coarse} concatenated the context with reference feature and refined the context through several convolutional layers. \cite{li2023neural} and \cite{qi2023motion} introduced offset diversity \cite{chan2021understanding} to obtain more accurate alignment for the largest-scale context. To reduce inaccurate alignment caused by large motions, \cite{qi2023motion} applied a self-attention-based context refinement module to the smallest-scale context. However, previous methods did not fully utilize the previously reconstructed signals at all scales. In addition, offset diversity method mainly focuses on local areas and lacks the ability to model the long-range correspondence.

As shown in Figure~\ref{context-enhance}, we propose a local-global context enhancement module to enhance context at each scale. For the local context enhancement, we adopt multi-scale deformable convolution to enhance the context at each scale in a progressive manner. Specifically, we first estimate the extra offsets $\tilde{o}^2_{t}$ of the smallest-scale feature $F^2_{t-1}$ by taking the $C^2_{t}$, $F^2_{t-1}$ and $\hat{v}^2_{t}$ as inputs. Then, based on $\hat{v}^2_{t}$ and $\tilde{o}^2_{t}$, we perform FGDC operation on $F^2_{t-1}$ to generate the enhanced context $\tilde{C}^2_{t}$.
Then, $\tilde{o}^2_{t}$ is upsampled and concatenated with $C^1_{t}$, $F^1_{t-1}$ and $\hat{v}^1_{t}$ to guide the offsets $\tilde{o}^1_{t}$ estimation of the next scale, forming a progressively guided manner. Based on the $\hat{v}^1_{t}$ and $\tilde{o}^1_{t}$, we perform FGDC on $F^1_{t-1}$ to generate the enhanced context $\tilde{C}^1_{t}$.
The generation of the enhanced largest-scale context $\tilde{C}^0_{t}$ is similar to other scale except that we use previously decoded offsets $\bar{o}_t$ instead of $\hat{v}_{t}$ as the base offsets to get better initialization.

To model the long-range correspondence, differently from \cite{qi2023motion}, we propose a cross-attention-based context enhancement module that enables the model to extract global information from reference feature. This module adopts a transformer-like architecture and is applied to the smallest-scale context to reduce computational cost. As shown in Figure~\ref{context-enhance} (b), context $C^2_{t}$ and reference feature $F^2_{t-1}$ are first normalized and projected to query (\textbf{Q}), key (\textbf{K}), and value (\textbf{V}). Then, the correlation between \textbf{Q} and \textbf{K} is calculated as an attention map (\textbf{A}). The projected \textbf{V} is multiplied by \textbf{A} to extract the global information. It is worth mentioning that a skip connection is used to stabilize training and convergence. Furthermore, we adopt a Gated-Dconv Feed-forward Network (GDFN) in \cite{zamir2022restormer} to enrich features with useful information. The global context enhancement module is repeated in 4 times in our implementation, finally generating the global enhanced context $\hat{C}^2_{t}$. 

\begin{figure}[!t]
  \centering
  \includegraphics[width=0.8\linewidth, trim = {230 175 235 175}, clip]{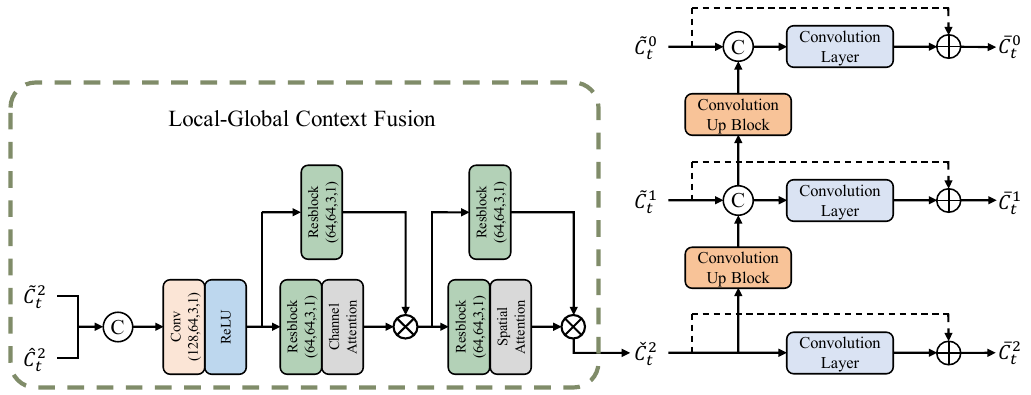}
  \vspace{-0.6cm}
  \caption{The network structure of multi-scale context fusion module.}
  \label{context-fusion}
  \vspace{-0.8cm}
\end{figure}

To fuse the local and global enhanced contexts $\tilde{C}^2_{t}$ and $\hat{C}^2_{t}$, as shown in Figure~\ref{context-fusion}, we use the channel-spatial attention mechanism from CBAM \cite{woo2018cbam} and redesign the submodules. Finally, we follow TCM \cite{sheng2022temporal} to fuse the local-global enhanced context $\check{C}^2_{t}$ with other scale contexts hierarchically and generate the final contexts $\bar C^0_t$, $\bar C^1_t$, $\bar C^2_t$.


\Section{EXPERIMENTS}
\label{sec:experiments}
\SubSection{Experimental Setup}
\label{sec:experimental_setup}
\textbf{Datasets.}
We use the Vimeo90K \cite{xue2019video} training set. During training, the videos are randomly cropped to 256 × 256 patches. For testing, we evaluate performance on multiple benchmark datasets including UVG \cite{mercat2020uvg}, MCL-JCV \cite{wang2016mcl}, HEVC \cite{sullivan2012overview} Class B, C, D, and E. The resolutions of the test datasets are from $416\times240$ to $1920\times1080$. 


\noindent
\textbf{Implementation and training Details.}
There is no public training code for TCM \cite{sheng2022temporal} and HEM \cite{li2022hybrid}. We use their released I-frame models and reproduce the P-frame models. For the HEM \cite{li2022hybrid} model, we found that multi-granularity quantization leads to training instability, so we reproduce it without multi-granularity quantization (denote as HEM$^*$). 
During training, the RD loss function is: $\mathcal{L} =R + \lambda D=R_{\hat v}+R_{\hat o}+R_{\hat f}+ \lambda D(x_t, \hat{x}_t)$. $R_{\hat v}$, $R_{\hat o}$ and $R_{\hat f}$ respectively denote the bitrate of the optical flow coding, the offset coding and the frame coding. $D(\cdot)$ denotes the distortion, which can  be $L_2$ loss or MS-SSIM. We adopt the same multi-stage training strategy as \cite{sheng2022temporal, li2022hybrid} and use 4 ${\lambda}$ values (MSE: 256, 512, 1024, 2048; MS-SSIM: 8, 16, 32, 64) to fit RD trade-off. We use the AdamW optimizer and set the batch size as 4. 


\begin{figure}[!t]
  \centering
  \vspace{-0.2cm}
  \includegraphics[width=\linewidth, trim = {55 155 80 170}, clip]{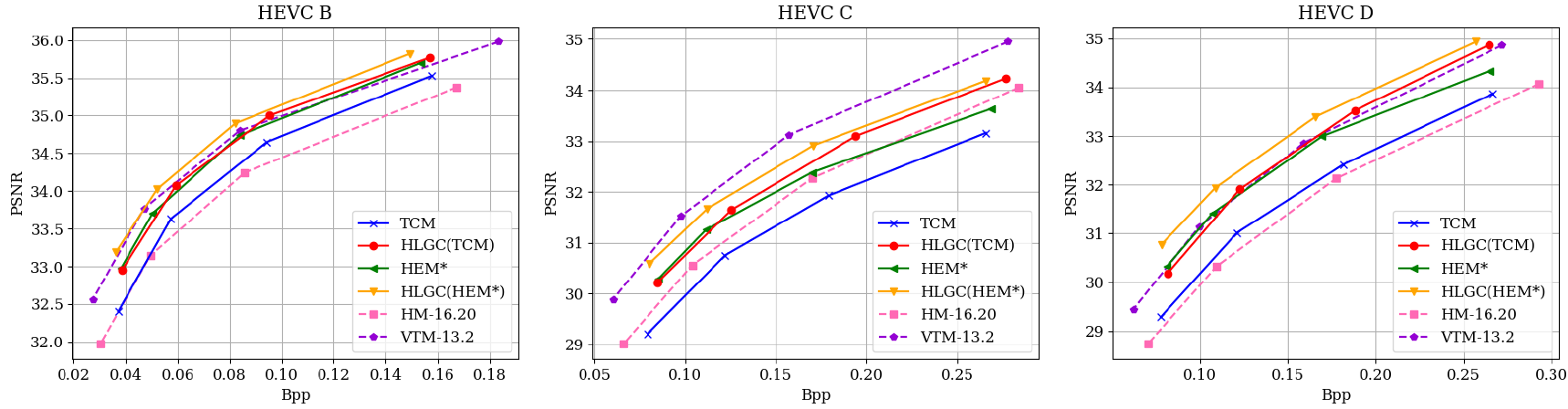}
  \vspace{-1.2cm}
  \caption{RD-curves on the HEVC B, C and D datasets.}
  \label{rd_curves}
  \vspace{-0.5cm}
\end{figure}

\begin{table}[!t]
\renewcommand{\arraystretch}{1.1}
\scriptsize
\centering
\vspace{-0.3cm}
\caption{BD-Rate (\%) comparison for PSNR. The anchor is HM-16.20.}\label{bdrate_psnr}
\vspace{-0.3cm}
\resizebox{\textwidth}{!}{
\begin{tabular}{c|cccccc|c}
    \toprule  
        \makebox[0.05\textwidth][c]{Methods} & HEVC B & HEVC C & HEVC D & HEVC E & UVG & MCL-JCV & Average \\
    \hline  
    TCM  &-8.3 &14.1 &-6.3 &11.9 &-12.4 &-7.9 &-1.5 \\ 
    HLGC(TCM)  &\textbf{-23.8} &\textbf{-10.6} &\textbf{-25.9} &\textbf{-5.2} &\textbf{-23.8} &\textbf{-16.1} &\textbf{-17.6} \\ 
    HEM*  &-23.5 &-5.7 &-24.7 &-19.6 &-26.3 &-16.9 &-19.4 \\ 
    HLGC(HEM*)  &\textbf{-32.1} &\textbf{-18.6} &\textbf{-34.6} &\textbf{-26.8} &\textbf{-32.9} &\textbf{-25.9} &\textbf{-28.5} \\ 
    VTM-13.2  &-29.1 &-28.4 &-26.5 &-33.2 &-26.3 &-30.1 &-28.9 \\ 
    \bottomrule 
\end{tabular}
}
\end{table}

\begin{table}[!t]
\renewcommand{\arraystretch}{1.1}
\scriptsize
\centering
\vspace{-0.3cm}
\caption{BD-Rate (\%) comparison for MS-SSIM. The anchor is HM-16.20.}\label{bdrate_ssim}
\vspace{-0.3cm}
\resizebox{\textwidth}{!}{
\begin{tabular}{c|cccccc|c}
    \toprule  
        \makebox[0.05\textwidth][c]{Methods} & HEVC B & HEVC C & HEVC D & HEVC E & UVG & MCL-JCV & Average \\
    \hline  
    TCM  &-49.0 &-42.4 &-52.5 &-24.5 &-25.4 &-37.1 &-38.4 \\ 
    HLGC(TCM)  &\textbf{-60.2} &\textbf{-52.6} &\textbf{-60.5} &\textbf{-56.3} &\textbf{-36.9} &\textbf{-47.6} &\textbf{-52.4} \\ 
    HEM*  &-59.2 &-53.6 &-61.4 &-56.8 &-36.1 &-45.7 &-52.1 \\ 
    HLGC(HEM*)  &\textbf{-60.9} &\textbf{-56.3} &\textbf{-64.5} &\textbf{-60.6} &\textbf{-40.8} &\textbf{-50.1} &\textbf{-55.5} \\ 
    VTM-13.2  &-28.7 &-28.2 &-27.2 &-28.3 &-22.6 &-30.2 &-28.1 \\ 
    \bottomrule 
\end{tabular}
}
\vspace{-0.5cm}
\end{table}
\SubSection{Experimental Results}
\label{sec:experimental_results}
To verify the effectiveness of our proposed method HLGC, we implement HLGC on the baselines TCM \cite{sheng2022temporal} and HEM* \cite{li2022hybrid}. Following the low delay encoding settings of the baselines, we set the intra period as 32 and test 96 frames for each video. We also compare with the traditional codecs HM-16.20 and VTM-13.2, which represent the best encoder of H.265 and H.266, respectively. Table~\ref{bdrate_psnr} and~\ref{bdrate_ssim} show the BD-Rate (\%) comparisons in terms of PSNR and MS-SSIM. The anchor is HM-16.20. The lower BD-Rate indicates better video compression performance. As we can see, our proposed method HLGC can significantly improve the performance of baselines TCM and HEM* on all test datasets. The performance improvement is particularly obvious on the HEVC C , D and E datasets, where our method achieves about 20.5$\%$ bitrate saving compared with TCM. When using TCM and HEM* as anchors, our HLGC method achieves average bitrate savings of 16.1{\%} and 9.1{\%} on all test datasets in terms of PSNR, respectively. As shown in Figure~\ref{rd_curves}, we also draw the RD-curves on the HEVC B, C and D datasets to verify the effectiveness of our method.


\SubSection{Ablation Study}
We conduct comprehensive ablation studies on TCM \cite{sheng2022temporal}. The comparisons are measured by BD-Rate (\%) for PSNR. Highlights are \textbf{\underline{best}}.
\label{sec:ablations}
\begin{table}[!t]
\renewcommand{\arraystretch}{1.1}
\scriptsize
\vspace{-0.5cm}
\caption{Ablation study on different compensation strategies.}\label{ablation_study_context_generation}
\vspace{-0.3cm}
\resizebox{\textwidth}{!}{
\begin{tabular}{cccc|cccccc|c}
    \toprule  
        \makebox[0.06\textwidth][c]{Methods} & \makebox[0.07\textwidth][c]{1/4 Scale} & \makebox[0.07\textwidth][c]{1/2 Scale} & \makebox[0.1\textwidth][c]{Original Scale} & B & C & D & E & UVG & MCL & Avg \\
    \hline  
    A  &Flow &Flow &Flow &0.0 &0.0 &0.0 &0.0 &0.0 &0.0 &0.0 \\ 
    B   &FGDC &Flow &Flow &10.4 &2.2 &0.2 &21.2 &11.1 &15.1 &10.0 \\ 
    C   &Flow &FGDC &Flow &-4.9 &-9.4 &-8.8 &-0.6 &-3.9 &\textbf{\underline{-2.2}} &-5.0 \\ 
    D   &Flow &Flow &FGDC &\textbf{\underline{-8.0}} &\textbf{\underline{-14.9}} &\textbf{\underline{-13.3}} &\textbf{\underline{-3.3}}  &\textbf{\underline{-4.3}} &-2.0 &\textbf{\underline{-7.6}} \\ 
    E   &Flow &FGDC &FGDC &-6.0 &-14.6 &-12.7 &4.9 &-2.8 &-1.5 &-5.5 \\ 
    F &DC &DC &DC &11.4 &1.2 &1.9 &17.6 &13.1 &25.3 &11.8 \\ 
    \bottomrule 
\end{tabular}
}
\vspace{-0.2cm}
\end{table}

\begin{table}[!t]
\renewcommand{\arraystretch}{1.1}
\scriptsize
\centering
\vspace{-0.2cm}
\caption{Ablation study on the local-global context enhancement module.}\label{ablation_local_global_context_enhancement}
\vspace{-0.3cm}
\resizebox{\textwidth}{!}{
\begin{tabular}{cccc|cccccc|c}
    \toprule  
        \makebox[0.06\textwidth][c]{Methods} & \makebox[0.06\textwidth][c]{1/4 Scale} & \makebox[0.06\textwidth][c]{1/2 Scale} & \makebox[0.1\textwidth][c]{Original Scale} & B & C & D & E & UVG & MCL & Avg \\
    \hline  
    D  &- &- &- &0.0 &0.0 &0.0 &0.0 &0.0 &0.0 &0.0 \\ 
    G   &FGDC &- &- &-0.3 &0.1 &0.2 &-5.1 &-4.3 &-2.0 &-1.9 \\ 
    H   &FGDC + CA &- &- &-0.6 &-0.9 &0.1 &-6.1 &-4.7 &-4.5 &-2.8 \\ 
    I   &FGDC + CA &FGDC &- &-2.6 &-1.9 &-2.9 &-5.4 &-5.2 &-5.1 &-3.9 \\ 
    J   &FGDC + CA &FGDC &FGDC &\textbf{\underline{-7.7}} &\textbf{\underline{-8.5}} &\textbf{\underline{-8.5}} &\textbf{\underline{-9.9}} &\textbf{\underline{-6.9}} 
    &\textbf{\underline{-5.8}}
    &\textbf{\underline{-7.9}} \\ 
    \bottomrule 
\end{tabular}
}
\vspace{-0.2cm}
\end{table}

\begin{table}[!t]
\renewcommand{\arraystretch}{1.15}
\scriptsize
\centering
\vspace{-0.2cm}
\caption{Model complexity comparison.}\label{complexity_comparison}
\vspace{-0.3cm}
\resizebox{0.82\textwidth}{!}{
\begin{tabular}{c|ccccc}
    \toprule  
     Methods& Parameters & FLOPs & MACs & Encoding Time & Decoding Time \\
    \hline  
    TCM  &10.71M &5.77T &2.88T &354ms &254ms \\ 
    HLGC(TCM)   &11.19M &7.19T &3.59T &576ms &436ms \\ 
    \bottomrule 
\end{tabular}
}
\vspace{-0.5cm}
\end{table}

\textbf{Different compensation strategies.}
As shown in Table~\ref{ablation_study_context_generation}, we apply flow-based warping (Flow) or flow-guided deformable compensation (FGDC) or deformable compensation (DC) to features at different scales. Both TCM \cite{sheng2022temporal} and HEM \cite{li2022hybrid} adopt method A (set as anchor) for motion compensation. Methods B, C, D and E compress extra deformable offsets for features at different scales respectively. Method F performs DC on features at all scales, which is used in \cite{yilmaz2023multi}. As we can see, compressing extra deformable offsets for 1/4 scale feature (method B) will cause significant performance degradation. This result shows that at 1/4 scale, the bitrate increase caused by compressing extra deformable offsets is larger than the prediction gain. When performing FGDC on larger-scale features (method C and D), we achieve bitrate savings compared to anchor method. And method D achieves better RD performance than method C, mainly because larger-scale features require finer reconstruction. We further find that performing FGDC on 1/2 and original scale simultaneously does not bring performance gains (method D and E). Comparison between method D and F proves that our hybrid context generation method is better than \cite{yilmaz2023multi}.


\textbf{Local-global context enhancement.}
To verify the effectiveness of the local-global context enhancement module, we conduct ablation studies in Table~\ref{ablation_local_global_context_enhancement}. We set method D as anchor and implement different context enhancement methods on it. It is shown that applying FGDC at 1/4 scale brings gains on datasets with small motions (HEVC E), but no gain is achieved on datasets with complex motions. When the cross-attention mechanism is additionally applied at 1/4 scale, the performance is improved on all test datasets. Compared with DCN focusing on local regions, the cross-attention (CA) mechanism additionally extracts global information and therefore achieves better RD performance. We apply cross-attention only at smallest-scale to save the computational cost. In addition, the bitrate saving is improved as the FGDC applied to more scales (method I and J). These comparative experiments demonstrate the effectiveness of our local-global context enhancement module.
\SubSection{Model Complexity}
\label{sec:computational_complexity}
In Table~\ref{complexity_comparison}, we compare the model complexity in parameters, FLOPs, MACs, encoding time and decoding time with basline method TCM\cite{sheng2022temporal}. The experiment is conducted on NVIDIA GeForce RTX 3090 GPU. We use one 1080p frame to measure complexity. For the encoding and decoding time, we report the model inference time on GPU. Comparing with baseline, our HLGC method slightly increases the model complexity (4.5\% extra parameters). Our encoding and decoding time is increased a little. However, in terms of PSNR, our HLGC method brings 16.1\% bitrate saving over the strong basline TCM\cite{sheng2022temporal}. We think this is a cost worth paying.


\Section{Conclusion}

In this paper, we propose a Hybrid Local-Global Context learning method to better generate high-quality contexts for neural video compression. For hybrid context generation, we combine the advantages of flow-based warping and deformable compensation in an optimal way. Our proposed method achieves better RD trade-off between compensation accuracy and bit cost for motion coding. Moreover, we design a local-global context enhancement module to further enhance the quality of contexts, which fully explore the local-global information of previous reconstructed signals. Experimental results on standard test datasets showed that our proposed HLGC method can significantly enhance the state-of-the-art methods.
\Section{Acknowledgment}
\sloppy{}

This work is financially supported by National Natural Science Foundation of China U21B2012 and  62072013, Shenzhen Science and Technology Program-Shenzhen Cultivation of Excellent Scientific and Technological Innovation Talents project (Grant No. RCJC20200714114435057), Shenzhen Science and Technology Program-Shenzhen Hong Kong joint funding project (Grant No. SGDX20211123144400001), this work is also financially supported for Outstanding Talents Training Fund in Shenzhen.

\Section{References}
\bibliographystyle{IEEEbib}
\bibliography{refs}

\end{document}